\documentclass[11pt,a4paper]{article}
\usepackage[T1]{fontenc}
\usepackage[utf8]{inputenc}
\usepackage[english]{babel}
\usepackage{lmodern}
\usepackage[margin=1in]{geometry}
\usepackage{amsmath,amssymb,amsthm,mathtools}
\usepackage{microtype}
\usepackage[nosort,nocompress]{cite}
\usepackage{algorithm,algpseudocode}
\usepackage[hidelinks]{hyperref}
\newtheorem{theorem}{Theorem}[section]
\newtheorem{lemma}[theorem]{Lemma}
\newtheorem{proposition}[theorem]{Proposition}

\title{Nearly Optimal Quantum Sample Complexity of Learning Fourier-Sparse Boolean Functions}
\author{Zekun Ye\\[0.5em]
  \small College of Computer and Data Science, Fuzhou University\\
  \small Fuzhou 350108, China\\
  \small \href{mailto:yezekun@fzu.edu.cn}{\texttt{yezekun@fzu.edu.cn}}}
\date{}
\hypersetup{
  pdftitle={Nearly Optimal Quantum Sample Complexity of Learning Fourier-Sparse Boolean Functions},
  pdfauthor={Zekun Ye},
  pdfsubject={Preprint},
  pdfkeywords={Quantum learning, exact learning, Fourier sparsity, quantum examples},
  bookmarksnumbered=true
}

\newcommand{\F}{\mathbb F}
\newcommand{\R}{\mathbb R}
\newcommand{\E}{\mathbb E}
\newcommand{\supp}{\operatorname{supp}}
\newcommand{\Maj}{\operatorname{Maj}}
\newcommand{\Var}{\operatorname{Var}}
\newcommand{\norm}[1]{\left\|#1\right\|}
\newcommand{\Sf}{\mathcal S_f}
\newcommand{\fail}{\mathsf{FAIL}}
\begin{document}
\maketitle
\begin{abstract}
We study the number of uniform quantum examples needed to exactly learn a Boolean function with at most $k$ nonzero Fourier coefficients. We give a learner with near-optimal $\widetilde O(k)$ sample complexity independent of the input length $n$. The learner uses only separate measurements of individual examples. This improves the previous $O(k^{3/2}\log^2 k)$ upper bound~\cite{ACLPW}. The main ingredient is a candidate set formed by symmetric differences of subsets from independent groups of Fourier samples. A multilinear majority polynomial combines unbiased random approximations into an accurate witness whose Fourier support lies in this set. Granularity then forces the set to contain the entire Fourier support, including subsets never directly sampled. For comparison, we derive the tight classical sample complexity $\Theta(nk)$ by combining known sparse-recovery and lower-bound results.
\end{abstract}

\section{Introduction}
\label{sec:intro}\label{page:main-start}
Learning from examples is a central task in computational learning theory~\cite{ADWSurvey}. Structure in an unknown function can make learning possible without examining its exponentially large domain. Fourier analysis describes such structure by expressing Boolean functions as linear combinations of parity functions. A function is \emph{Fourier-sparse} when only a few coefficients are nonzero.

Fourier sparsity counts the nonzero terms in the representation and provides a natural parameter for relating representation complexity to the amount of data needed for learning. The class includes signed parity functions, which have sparsity one, and every function depending on $d$ variables, whose sparsity is at most $2^d$. Unlike dependence on a small set of variables, Fourier sparsity also permits the relevant parity functions to involve arbitrarily many input coordinates~\cite{ACLPW}.

\paragraph*{Exact Learning.}
In exact learning, the goal is to identify an unknown target from a known class of functions~\cite{ADWSurvey,ACLPW}. A learner must output a classical description that agrees with the target on every input, with high success probability. For Fourier-sparse Boolean functions, we recover the complete Fourier representation. This requires identifying every nonzero spectral component, including those that contribute little to average prediction error.

\paragraph*{Classical and Quantum Examples.}
We study $f:\F_2^n\to\{-1,1\}$ under the uniform distribution. A uniform classical example is a labeled pair $(X,f(X))$, where $X$ is uniformly distributed on $\F_2^n$. A uniform quantum example is the state
\begin{equation}
 |\psi_f\rangle\coloneqq2^{-n/2}\sum_{x\in\F_2^n}|x\rangle|b_f(x)\rangle,
 \qquad b_f(x)\coloneqq\frac{1-f(x)}2.
 \label{eq:example}
\end{equation}
This quantum-example model was introduced by Bshouty and Jackson~\cite{BJ}; see~\cite{ADWSurvey} for a broader account. The objective is to output an exact classical description of the target using as few examples as possible.

Following~\cite{HR}, the \emph{sample complexity} of exact learning is the minimum number of independent uniform examples needed by a learner to identify any target in the class, with success probability at least $2/3$. Classical and quantum sample complexity refer to the corresponding types of examples~\cite{ACLPW}.

The essential quantum operation is \emph{Fourier sampling}. Conditional on an event of probability $1/2$, one copy yields a subset of variables $S\subseteq[n]$ with probability $\widehat f(S)^2$~\cite{BJ,ACLPW}. For a signed parity function, a successful Fourier sample reveals $S$, and one further classical example determines its sign. Thus learning a signed parity function requires only a constant number of quantum examples, compared with $\Theta(n)$ uniform classical examples~\cite{ACLPW}, in both cases with constant success probability. Fourier-sparse functions extend this basic separation from one parity function to a sparse combination of parity functions. Determining their sample complexity asks how much of the advantage survives when several spectral components must be identified simultaneously.

\paragraph*{The Gap in Known Bounds.}
Arunachalam, Chakraborty, Lee, Paraashar, and de Wolf \cite{ACLPW} gave an $O(k^{3/2}\log^2 k)$ quantum upper bound independent of $n$, together with an $\Omega(k\log k)$ lower bound for $k\le2^{cn}$ and any fixed $c\in(0,1)$. They conjectured that \textbf{$O(k\,\mathrm{polylog}\,k)$ uniform quantum examples suffice to learn every $k$-Fourier-sparse Boolean function exactly.}

\paragraph*{Why the General Case Is Difficult.}
For $k\ge2$, if every nonzero Fourier coefficient has magnitude comparable to $1/\sqrt{k}$, coupon collection finds the support from $O(k\log k)$ Fourier samples. Fixed-target recovery then determines the coefficients from $O(k\log k)$ classical examples~\cite{CP}. In general, however, a nonzero coefficient can have magnitude of order $1/k$, so its corresponding subset may appear with probability of order $1/k^2$~\cite{GOSSW,ACLPW}. Waiting to observe every support element can therefore be too expensive. Existing learners instead find the linear span of the support, but recovering within that span incurs an additional polynomial cost in $k$~\cite{Sanyal,ACLPW}. The challenge is to identify a sufficiently small candidate set containing the support without observing every support element directly.

\subsection{Main Results}
For a Boolean function $f:\F_2^n\to\{-1,1\}$ and a subset $S\subseteq[n]=\{1,\ldots,n\}$, the Fourier coefficient of $f$ on $S$ is
\[
 \widehat f(S)\coloneqq2^{-n}\sum_{x\in\F_2^n}f(x)(-1)^{\sum_{i\in S}x_i}.
\]
We call $f$ \emph{$k$-Fourier-sparse} if at most $k$ of these coefficients are nonzero.

Our main result is a near-optimal $\widetilde O(k)$ upper bound, establishing the conjectured near-linear sample bound~\cite{ACLPW}. To state it precisely, put $\Lambda(k)\coloneqq\log^2 k/\log\log k$ for $k\ge3$ and $\Lambda(1)=\Lambda(2)=1$. Both results below hold for all integers $n\ge1$ and $1\le k\le2^n$.

Throughout, $\log$ is base two and $\ln$ is natural. Logarithmic factors in asymptotic bounds are understood to be at least one. The notation $\widetilde O(k)$ suppresses factors polynomial in $\log k$.

\begin{theorem}[Quantum upper bound]
\label{thm:main}
There exists a quantum learner that exactly learns any $k$-Fourier-sparse Boolean function $f:\F_2^n\to\{-1,1\}$ using $O(k\Lambda(k))$ uniform quantum examples, with success probability at least $2/3$. The learner reconstructs the following Fourier list exactly:
\[
 \{(S,\widehat f(S)):S\subseteq[n],\ \widehat f(S)\ne0\}.
\]
The learner measures each copy separately. For any $\delta\in(0,1/3)$, independent repetition uses $O(k\Lambda(k)\log(1/\delta))$ copies to recover the list exactly, with success probability at least $1-\delta$.
\end{theorem}

The earlier learner~\cite{ACLPW} first identifies the linear span of the Fourier support and then recovers the function within that span. Its sample bound is $O(k(d+1)\log k)$, where $d$ is the span dimension; substituting $d=O(\sqrt{k}\log k)$~\cite{Sanyal} gives the $k^{3/2}$ dependence. Our method instead constructs a candidate set $C$ with $\log|C|=O(\Lambda(k))$ using $O(k\Lambda(k))$ examples. Recovery then uses $O(k\log|C|)$ further examples~\cite{CP}, giving the near-linear total.

Theorem~\ref{thm:main} is near-optimal: the $\Omega(k\log k)$ lower bound of~\cite{ACLPW} applies when $k\le2^{cn}$ for fixed $c\in(0,1)$, and the same bound for all $k$ follows from the quantum coupon-collector theorem~\cite{QCC}. Appendix~\ref{app:quantum-lower} records this consequence and its reduction. Thus the gap is at most $O(\log k/\log\log k)$ throughout the parameter range.

For comparison, combining known recovery and lower-bound results gives the following tight classical characterization.

\begin{theorem}[Tight classical sample complexity]
\label{thm:classical}
The number of independent uniform classical examples necessary and sufficient to learn the class of $k$-Fourier-sparse Boolean functions on $\F_2^n$ exactly is $\Theta(nk)$, with success probability at least $2/3$.
\end{theorem}

Haviv and Regev~\cite[Corollary~3.6]{HR} obtained an $O(nk\log k)$ upper bound with failure probability $2^{-\Omega(n\log k)}$. Theorem~\ref{thm:classical} removes a factor of $\log k$ when only constant error is required; it does not assert the same improvement at their exponentially small error.

\subsection{Related Work}
\paragraph*{Fourier Methods and Learning Models.}
Fourier methods have long been central to learning Boolean functions. Linial, Mansour, and Nisan~\cite{LMN} used Fourier concentration to learn constant-depth circuits. Kushilevitz and Mansour~\cite{KM} gave a membership-query algorithm for learning decision trees; in that model, the learner chooses which inputs to label. Our problem concerns exact learning from independent uniform examples. The surveys of Arunachalam and de Wolf~\cite{ADWSurvey} and Chatterjee~\cite{Menagerie} review quantum learning models and their sample complexities.

\paragraph*{Quantum Learning.}
Bernstein and Vazirani~\cite{BV} showed how to identify a parity function with one quantum query. In the quantum-example model, Bshouty and Jackson~\cite{BJ} gave an algorithm for learning disjunctive normal form (DNF) formulas under the uniform distribution. At{\i}c{\i} and Servedio~\cite{ASJuntas} subsequently developed algorithms for learning and testing juntas from uniform classical and quantum examples, with sample bounds independent of the input length. For distribution-independent probably approximately correct (PAC) and agnostic learning, Arunachalam and de Wolf~\cite{ADWOptimal} proved that quantum and classical sample complexities agree up to constant factors. Our setting fixes the uniform distribution and requires exact identification of the target.

\paragraph*{Exact Learning of Fourier-Sparse Functions.}
For the problem studied here, Arunachalam, Chakraborty, Lee, Paraashar, and de Wolf~\cite{ACLPW} proved an $O(k^{3/2}\log^2 k)$ quantum upper bound and an $\Omega(k\log k)$ quantum lower bound for $k\le2^{cn}$ and any fixed $c\in(0,1)$. Theorem~\ref{thm:main} improves the quantum upper bound to $\widetilde O(k)$. For classical examples, Haviv and Regev~\cite{HR} proved an $O(nk\log k)$ upper bound and an $\Omega(k(n-\log k))$ lower bound. By combining known results, we obtain the tight classical sample complexity $\Theta(nk)$ at constant error for all $1\le k\le2^n$ (Theorem~\ref{thm:classical}).

\paragraph*{Sparse Recovery.}
Exact recovery from partial Fourier information was developed by Cand\`es, Romberg, and Tao~\cite{CRT}. Cand\`es and Plan~\cite{CP} established exact recovery of each fixed sparse vector under general random measurement models. Rudelson and Vershynin~\cite{RV} and Haviv and Regev~\cite{HRRIP} studied uniform recovery, in which one measurement matrix must recover all sparse vectors simultaneously. These works develop sparse-recovery techniques relevant to exact learning.

\subsection{Proof Overview}
The quantum upper-bound proof has two parts: first construct a candidate set containing the entire Fourier support, and then exactly learn the coefficients within that set. These parts use independent batches of examples. We describe the main ideas here and defer the quantitative estimates to Sections~\ref{sec:coverage}--\ref{sec:algorithm}.

\paragraph*{Part I: Covering the Fourier Support.}
The algorithm collects several independent groups of Fourier samples. It forms a candidate set $C$ by selecting at most one observed subset from each group and taking their symmetric difference. The resulting set can contain subsets never observed directly, which is essential for recovering coefficients that Fourier sampling rarely reveals.

To prove coverage, we introduce a weighted average of the sampled parity functions in each group. A subset $S$ is sampled with probability $\widehat f(S)^2$~\cite{BJ,ACLPW}, so assigning the corresponding parity function the weight $1/\widehat f(S)$ gives expected contribution $\widehat f(S)\chi_S(x)$. Summing over the possible outcomes recovers $f(x)$ in expectation. Averaging more samples within a group decreases the variance. These weighted functions are only tools for the analysis: their weights are unknown, and the learner never computes them. The candidate set itself is formed solely from the observed subsets.

We combine the independent averages using the multilinear extension of majority. At an input where all signs agree, changing a minority of signs cannot change the majority. As a result, the expansion of this polynomial about the correct Boolean value has no low-degree error terms. Every surviving term contains many independent, mean-zero errors. Their mixed terms vanish in the second moment, making the expected squared error small. Multilinearity also ensures that every parity function in the resulting auxiliary function is indexed by a subset in the candidate set.

Finally, Boolean Fourier granularity~\cite{GOSSW} supplies a minimum possible size of a nonzero coefficient. By Parseval's identity, omitting any subset in the true support would force the auxiliary function to incur a definite squared error. The small expected error therefore implies that the candidate set contains the entire Fourier support with high probability. This proves support containment without separately estimating the Fourier coefficients.

\paragraph*{Part II: Recovering the Fourier Coefficients.}
Let $C$ be the candidate set constructed in Part~I. Once $C$ is fixed and contains the support, the unknown Fourier coefficients form a sparse vector in the dictionary $\{\chi_S:S\in C\}$. Uniform classical examples provide exact linear equations for this vector. Orthogonality of parity functions verifies the hypotheses of the Cand\`es--Plan recovery theorem~\cite{CP}. The true coefficient vector is the unique solution of $\ell_1$ minimization with high probability. These examples are independent of those used to construct the candidate set. Conditional on that construction, the theorem applies to a fixed dictionary and a fixed target, requiring $O(k\log|C|)$ examples.

Applied to all $2^n$ parity functions, the same recovery theorem~\cite{CP} also gives the $O(nk)$ classical upper bound in Theorem~\ref{thm:classical}.

\section{Preliminaries}
\label{sec:prelim}
We write $\F_2$ for the field with two elements and $\R$ for the real numbers. For a positive integer $a$, let $[a]\coloneqq\{1,\ldots,a\}$. The symbols $\Pr$ and $\E$ denote probability and expectation. For a real random variable $Z$ with finite second moment,
\[
 \Var(Z)\coloneqq\E[(Z-\E Z)^2]=\E[Z^2]-(\E Z)^2.
\]
A subscript specifies the random variable or distribution being averaged over. In particular, $\E_x$ averages over a uniform input, whereas $\E_{\mathrm{samples}}$ averages over the samples used by the learner. Without a subscript, the source of randomness is the one specified locally. We write $\mathbf1_A$ for the indicator of an event or condition $A$; an empty product equals one.

For real vectors, $\langle u,v\rangle\coloneqq\sum_i u_i v_i$ and $\|u\|_1\coloneqq\sum_i|u_i|$. The notation $I_M$ denotes the $M$-dimensional identity matrix, and $A^\top$ denotes transpose. The space $\R^C$ consists of real vectors indexed by a finite set $C$. A vector is $s$-sparse if at most $s$ entries are nonzero.

\subsection{Fourier Analysis}
Identify every subset $S\subseteq[n]$ with its incidence vector in $\F_2^n$, and let $S\cdot x$ denote the inner product modulo two. The same notation is used for binary vectors of any equal length. We use $+$ for addition modulo two. For subsets, $S+T=(S\setminus T)\cup(T\setminus S)$ is their symmetric difference. This convention applies to subsets of any coordinate set. For every real function $h:\F_2^n\to\R$, define
\[
 \chi_S(x)\coloneqq(-1)^{S\cdot x},\qquad
 \widehat h(S)\coloneqq\E_x[h(x)\chi_S(x)],\qquad
 \norm{h}_2^2\coloneqq\E_x[h(x)^2].
\]
The function $\chi_S$ is the $\{-1,1\}$-valued \emph{parity function} on $S$. The subset $S$ specifies its variables and is represented by its $n$-bit incidence vector.
Unless otherwise specified, $x$ is uniform on $\F_2^n$. The Fourier support of $h$ is $\supp(\widehat h)\coloneqq\{S\subseteq[n]:\widehat h(S)\ne0\}$. For the target $f$, define its Fourier support and sparsity by
\[
 \Sf\coloneqq\{S\subseteq[n]:\widehat f(S)\ne0\},\qquad
 s\coloneqq|\Sf|.
\]
The promise on the target is $s\le k$. For any $S,T\subseteq[n]$, the parity functions satisfy
\[
 \chi_S\chi_T=\chi_{S+T},\qquad
 \E_x[\chi_S(x)\chi_T(x)]=\mathbf1_{S=T}.
\]
They form an orthonormal basis under the uniform inner product, giving Fourier inversion and Parseval's identity~\cite{ODonnell}:
\begin{equation}
 h(x)=\sum_{S\subseteq[n]}\widehat h(S)\chi_S(x),\qquad
 \norm{h}_2^2=\sum_{S\subseteq[n]}\widehat h(S)^2.
 \label{eq:fourier}
\end{equation}
The second equality is \emph{Parseval's identity}: the average squared value of a function equals the sum of the squares of its Fourier coefficients. It follows by expanding the square in the first equality; orthogonality makes every mixed term vanish. In particular, $\sum_S\widehat f(S)^2=1$ for Boolean $f$.

A function has Fourier sparsity one if and only if it is a signed parity function $f=\sigma\chi_S$, where $S\subseteq[n]$ and $\sigma\in\{-1,1\}$. Indeed, Parseval forces its sole nonzero coefficient to have magnitude one. The case $S=\varnothing$ gives the two constant functions.

\begin{lemma}[Consequence of granularity~{\cite[Theorem~8.1]{GOSSW}}]
\label{lem:granularity}
Let $k\ge1$ be an integer and let $f:\F_2^n\to\{-1,1\}$ have Fourier sparsity at most $k$. Every nonzero Fourier coefficient satisfies
\begin{equation}
 S\in\Sf\quad\Longrightarrow\quad |\widehat f(S)|\ge\frac1k.
 \label{eq:granularity}
\end{equation}
\end{lemma}

\subsection{Quantum Examples}
We use the uniform quantum-example state defined in~\eqref{eq:example}. The ket $|x\rangle$ is the computational-basis vector labeled by $x$; adjacent kets and $\otimes$ denote tensor products. The single-qubit Hadamard gate and the uniform qubit state are
\[
 H\coloneqq\frac1{\sqrt2}\begin{pmatrix}1&1\\1&-1\end{pmatrix},
 \qquad |+\rangle\coloneqq\frac{|0\rangle+|1\rangle}{\sqrt2}.
\]

\begin{lemma}[Measurements of a quantum example~\cite{BJ,ACLPW}]
\label{lem:sampling}
Computational-basis measurement of one copy of~\eqref{eq:example} gives a uniform classical example. Alternatively, apply $H$ to every qubit and measure in the computational basis, declaring success when the label qubit is $1$. This succeeds with probability $1/2$ and, conditional on success, returns each subset $S\subseteq[n]$ with probability $\widehat f(S)^2$.
\end{lemma}

We call a successful outcome in Lemma~\ref{lem:sampling} a Fourier sample. All Fourier samples used below come from different copies. The learner has no membership-query access or access to an inverse state-preparation circuit. Both Fourier measurements and classical-example measurements are charged to the copy budget.

\subsection{Probability Inequalities}
\begin{proposition}[Markov's inequality]
\label{prop:markov}
Let $Z$ be a nonnegative random variable with finite expectation. For every $a>0$,
\begin{equation}
 \Pr[Z\ge a]\le\frac{\E Z}{a}.
 \label{eq:markov}
\end{equation}
\end{proposition}

\begin{proposition}[Chernoff--Hoeffding bound~\cite{Hoeffding}]
\label{prop:chernoff}
Let $Z_1,\ldots,Z_R$ be independent random variables taking values in $\{0,1\}$ and put $\mu\coloneqq\sum_{i=1}^R\E Z_i$. For $0<\eta<1$,
\[
 \Pr\!\left[\sum_{i=1}^R Z_i\le(1-\eta)\mu\right]
 \le e^{-\eta^2\mu/2}.
\]
\end{proposition}

\begin{proposition}[Hoeffding's inequality~\cite{Hoeffding}]
\label{prop:hoeffding}
Let $Z_1,\ldots,Z_R$ be independent random variables taking values in $[0,1]$, and set $\mu\coloneqq\sum_{i=1}^R\E Z_i$. For every $t>0$,
\[
 \Pr\!\left[\sum_{i=1}^R Z_i-\mu\ge t\right]
 \le \exp\!\left(-\frac{2t^2}{R}\right).
\]
\end{proposition}

\section{Covering the Fourier Support}
\label{sec:coverage}
The construction has two sampling choices: the number of Fourier samples in each group and the number of independent groups. Let $r\ge16$ be the \emph{sampling ratio}: each group contains $rk$ Fourier samples, rounded up to an integer. Thus $r$ measures the group size relative to the sparsity bound $k$. We set
\begin{equation}
 m\coloneqq\lceil rk\rceil,\qquad
 g\coloneqq2\left\lceil\frac{\ln(100k^2)}{\ln(r/4)}\right\rceil+1.
 \label{eq:parameters}
\end{equation}
The construction uses $gm$ successful Fourier samples in total. The group count $g$ is odd because the groups are combined by a majority polynomial with $g$ arguments. At least $(g+1)/2$ inputs must change to reverse a unanimous majority; this threshold controls the error bound below.

For each $i\in[g]$, draw $m$ independent Fourier samples $S_{i1},\ldots,S_{i,m}$, independently across groups. Put $T_i\coloneqq\{S_{i1},\ldots,S_{i,m}\}$ and define
\begin{equation}
 C\coloneqq(T_1\cup\{0\})+\cdots+(T_g\cup\{0\}).
 \label{eq:candidates}
\end{equation}
Here the sumset consists of all symmetric differences obtained by choosing one subset from each $T_i\cup\{0\}$, where $0$ denotes the empty subset. Because every Fourier sample belongs to $\Sf$, we have $|T_i|\le s\le k$. Consequently,
\begin{equation}
 1\le |C|\le\prod_{i=1}^g(|T_i|+1)\le(k+1)^g.
 \label{eq:candidate-size}
\end{equation}
The candidate set may contain subsets outside $\Sf$; the required property is $\Sf\subseteq C$.

\begin{theorem}[Coverage of the complete support]
\label{thm:coverage}
For the set in~\eqref{eq:candidates}, with the parameters in~\eqref{eq:parameters},
\[
 \Pr[\Sf\nsubseteq C]\le\frac1{50r}.
\]
\end{theorem}

We prove Theorem~\ref{thm:coverage} by constructing an auxiliary function whose Fourier support lies in $C$. Each sample group gives an unbiased approximation to $f$ with variance at most $1/r$. The multilinear extension of majority combines these approximations into a function with small expected squared error. Multilinearity ensures that its Fourier support remains in $C$. If $C$ omits a nonzero Fourier coefficient of $f$, granularity and Parseval's identity force the squared error to be at least $1/k^2$. Markov's inequality then gives the claimed probability bound.

\subsection{Independent Unbiased Approximations}
For each group, introduce the random real function
\begin{equation}
 P_i(x)\coloneqq\frac1{m}\sum_{j=1}^{m}
           \frac{\chi_{S_{ij}}(x)}{\widehat f(S_{ij})}.
 \label{eq:group-mean}
\end{equation}
The denominators are nonzero because Fourier samples lie in $\Sf$. These functions are used solely in the proof; they are not computed by the learner. In particular, the candidate set in~\eqref{eq:candidates} does not require their coefficients.

\begin{lemma}[Moments of a group mean]
\label{lem:moments}
For every fixed $x\in\F_2^n$ and every $i\in[g]$,
\[
 \E P_i(x)=f(x),\qquad
 \Var(P_i(x))\le\frac1{r}.
\]
For fixed $x$, the random variables $P_1(x),\ldots,P_g(x)$ are independent.
\end{lemma}
\begin{proof}
Let $S$ be one Fourier sample. Its distribution is $\Pr[S=U]=\widehat f(U)^2$, so
\begin{align*}
 \E\left[\frac{\chi_S(x)}{\widehat f(S)}\right]
 &=\sum_{U\in\Sf}\widehat f(U)^2
                  \frac{\chi_U(x)}{\widehat f(U)}
 =\sum_{U\in\Sf}\widehat f(U)\chi_U(x)=f(x),\\
 \E\left[\left(\frac{\chi_S(x)}{\widehat f(S)}\right)^2\right]
 &=\sum_{U\in\Sf}\widehat f(U)^2\frac1{\widehat f(U)^2}=s.
\end{align*}
Since $f(x)^2=1$, one summand has variance $s-1$. The $m$ summands are independent, so their average satisfies
\[
 \Var(P_i(x))=\frac{s-1}{m}\le\frac{k}{m}\le\frac1r.
\]
Independence of the groups proves the final assertion.
\end{proof}

Only independence across groups at a fixed $x$ is used. Values at different inputs may be correlated; the proof later averages over $x$ and does not require a union bound over the input domain or a pointwise approximation guarantee for every group.

\subsection{The Multilinear Extension of Majority}
For an odd number $g$ of groups, let $\Maj(u)\in\{-1,1\}$ be the majority of $u\in\{-1,1\}^g$. Its multilinear extension to $\R^g$ is
\begin{equation}
 M_g(z_1,\ldots,z_g)
  \coloneqq\sum_{u\in\{-1,1\}^g}\Maj(u)
                  \prod_{i=1}^g\frac{1+u_i z_i}{2}.
 \label{eq:majority-extension}
\end{equation}
At a Boolean vector $z$, all summands vanish except the one with $u=z$, which has value $\Maj(z)$. Thus this formula is the unique multilinear polynomial agreeing with majority on the Boolean cube. It remains a polynomial identity for arbitrary real $z_i$. No probabilistic interpretation of the interpolation factors is used when $|z_i|>1$.

\begin{lemma}[Vanishing of low-degree terms]
\label{lem:vanishing}
For each $y\in\{-1,1\}$ there are coefficients $c_J^{(y)}$ with $|c_J^{(y)}|\le1$ such that, for every $(e_1,\ldots,e_g)\in\R^g$,
\begin{equation}
 M_g(y+e_1,\ldots,y+e_g)-y
 =\sum_{\substack{J\subseteq[g]\\|J|\ge(g+1)/2}}
       c_J^{(y)}\prod_{i\in J}e_i.
 \label{eq:vanishing}
\end{equation}
\end{lemma}
\begin{proof}
The translated polynomial is multilinear and its constant term is $M_g(y,\ldots,y)=y$. For a nonempty $J\subseteq[g]$, fix every coordinate outside $J$ to $y$. Write $J^c\coloneqq[g]\setminus J$, and let
$R_J(z_J)\coloneqq M_g(z_J,y_{J^c})$ denote this restricted polynomial.
The coefficient of $\prod_{i\in J}e_i$ after substituting $z_i=y+e_i$ equals the coefficient of $\prod_{i\in J}z_i$ in $R_J$: a monomial missing a variable cannot contribute to the full product. Multiplying $R_J(u)$ by $\prod_{i\in J}u_i$ and averaging over uniform $u\in\{-1,1\}^J$ removes every other monomial, so this coefficient is
\[
 c_J^{(y)}=2^{-|J|}\sum_{u\in\{-1,1\}^J}
                 \Maj(u,y_{J^c})\prod_{i\in J}u_i.
\]
It is an average of signs, so $|c_J^{(y)}|\le1$. If $|J|\le(g-1)/2$, the fixed coordinates already determine the majority, and the restricted function is constant. Its coefficient on the nonempty product $\prod_{i\in J}u_i$ is therefore zero. This proves the expansion, as a polynomial identity for arbitrary real errors.
\end{proof}

\subsection{Mean-Square Error and Support Coverage}
To bound the approximation error and then establish support containment, define
\[
 F^\star(x)\coloneqq M_g(P_1(x),\ldots,P_g(x)).
\]
We first bound its approximation error, and then determine which subsets can appear in its Fourier representation.

\begin{lemma}[Mean-square error bound]
\label{lem:ms}
The random function $F^\star$ satisfies
\begin{equation}
 \E_{\mathrm{samples}}\norm{F^\star-f}_2^2
 \le\frac2r\left(\frac4r\right)^{(g-1)/2}.
 \label{eq:ms}
\end{equation}
\end{lemma}
\begin{proof}
Fix $x$, and put $y\coloneqq f(x)$ and $e_i\coloneqq P_i(x)-f(x)$ for $i\in[g]$. The expectations below are over the sample groups, with $x$ fixed. By Lemma~\ref{lem:moments}, the $e_i$ are independent, $\E e_i=0$, and $\E e_i^2\le1/r$. Set
\[
 \mathcal I\coloneqq\{J\subseteq[g]:|J|\ge(g+1)/2\},
 \qquad e_J\coloneqq\prod_{i\in J}e_i.
\]
Squaring~\eqref{eq:vanishing} gives
\[
 (F^\star(x)-f(x))^2
 =\sum_{J,K\in\mathcal I}c_J^{(y)}c_K^{(y)}e_Je_K.
\]
For distinct $J,K$, independence gives
\[
 \E[e_Je_K]
 =\prod_{i\in J\cap K}\E[e_i^2]
  \prod_{i\in J+K}\E[e_i]=0,
\]
because $J+K$ is nonempty. For $J=K$, independence gives
\[
 \E[e_J^2]=\prod_{i\in J}\E[e_i^2]\le r^{-|J|}.
\]
Consequently, using $|c_J^{(y)}|\le1$ and $r\ge1$,
\begin{align*}
 \E[(F^\star(x)-f(x))^2]
 &=\sum_{J\in\mathcal I}(c_J^{(y)})^2\E[e_J^2]\\
 &\le\sum_{j=(g+1)/2}^g\binom gj r^{-j}
 \le 2^g r^{-(g+1)/2}
 =\frac2r\left(\frac4r\right)^{(g-1)/2}.
\end{align*}
Averaging over the uniform input $x$ proves~\eqref{eq:ms}. All sums are finite, so the order of the two expectations can be exchanged; no union bound over inputs is needed.
\end{proof}

\begin{lemma}[Fourier support of the analysis function]
\label{lem:support}
For every realization of the samples, $\supp(\widehat{F^\star})\subseteq C$.
\end{lemma}
\begin{proof}
Each $P_i$ in~\eqref{eq:group-mean} is a linear combination of parity functions indexed by subsets in $T_i$. Use the original multilinear expansion
\[
 M_g(z)=\sum_{J\subseteq[g]}\gamma_J\prod_{i\in J}z_i,
\]
where the real coefficients $\gamma_J$ depend only on majority. In a term $\prod_{i\in J}P_i(x)$, expanding the product selects one parity function $\chi_{U_i}(x)$ with $U_i\in T_i$ from each $i\in J$. Their product equals $\chi_{\sum_{i\in J}U_i}(x)$. Taking $U_i=0$ for $i\notin J$ shows that this sum is in $C$. The empty product contributes only the empty subset, also in $C$. Adding these terms can cancel coefficients but cannot introduce subsets outside $C$.
\end{proof}

\begin{proof}[Proof of Theorem~\ref{thm:coverage}]
If $\Sf\nsubseteq C$, choose $S\in\Sf\setminus C$. Lemma~\ref{lem:support} gives $\widehat{F^\star}(S)=0$. By Parseval and Lemma~\ref{lem:granularity},
\[
 \norm{F^\star-f}_2^2
 =\sum_U(\widehat{F^\star}(U)-\widehat f(U))^2
 \ge\widehat f(S)^2\ge\frac1{k^2}.
\]
This implication holds for each realization. Apply Markov's inequality (Proposition~\ref{prop:markov}) to the nonnegative random variable $\|F^\star-f\|_2^2$ with threshold $1/k^2$. Together with Lemma~\ref{lem:ms}, this gives
\begin{align*}
 \Pr[\Sf\nsubseteq C]
 &\le k^2\E\norm{F^\star-f}_2^2\\
 &\le\frac{2k^2}{r}\left(\frac4r\right)^{(g-1)/2}
 \le\frac{2k^2}{r}\frac1{100k^2}
 =\frac1{50r}.
\end{align*}
The penultimate inequality follows from the choice of the group count $g$ in~\eqref{eq:parameters} and $r/4>1$.
\end{proof}

\section{Exact Recovery from a Candidate Set}
\label{sec:recovery}
Once a candidate set contains the Fourier support, the problem becomes sparse recovery in a known finite dictionary of parity functions. We first state the external recovery result in the form we use.

\begin{lemma}[Fixed-vector recovery, Cand\`es--Plan~{\cite[Theorem~1.1]{CP}}]
\label{lem:cp}
Let $M\ge2$, and let $v$ be a random vector in $\R^M$ such that $\E[vv^\top]=I_M$ and $\max_i|v_i|^2\le\mu$ almost surely. Fix an $s$-sparse vector $w\in\R^M$ and a number $\beta>0$. There is a universal constant $c$ such that, for $q$ independent copies $v_1,\ldots,v_q$ of $v$ with
\[
 q\ge c(1+\beta)\mu s\ln M,
\]
the vector $w$ is the unique minimizer of
\[
 \min_{z\in\R^M}\norm z_1
 \quad\text{subject to}\quad
 \langle v_j,z\rangle=\langle v_j,w\rangle\quad(1\le j\le q)
\]
with probability at least $1-5/M-e^{-\beta}$.
\end{lemma}

Fix a constant $c$ for which Lemma~\ref{lem:cp} holds. Define the recovery budget uniformly for all $M\ge1$ by
\begin{equation}
 q(k,M)\coloneqq
 \left\lceil c(1+\ln20)k\ln\bigl(\max\{M,100\}\bigr)\right\rceil.
 \label{eq:recovery-budget}
\end{equation}

The constant cutoff permits a reduction of the small-dimensional case to dimension $100$, without a separate recovery algorithm.

\begin{lemma}[Recovery on a known candidate set]
\label{lem:recovery}
Let $C\subseteq\F_2^n$ contain $M\ge1$ distinct subsets, and suppose $\Sf\subseteq C$. From $q(k,M)=O(k\log M)$ independent uniform classical examples, Algorithm~\ref{alg:recovery} recovers all Fourier coefficients of $f$ with probability at least $0.9$. This statement also holds conditional on any independently generated candidate set that contains $\Sf$.
\end{lemma}
\begin{proof}
Fix a nonempty $C$ containing $\Sf$, and put
\[
 M\coloneqq|C|,\qquad
 v(X)\coloneqq(\chi_S(X))_{S\in C},\qquad
 w\coloneqq(\widehat f(S))_{S\in C}.
\]
For uniform $X$, orthogonality of parity functions gives
\[
 \E[v(X)_S v(X)_T]=\mathbf1_{S=T}\quad(S,T\in C),
 \qquad \E[v(X)v(X)^\top]=I_M.
\]
Every entry of $v(X)$ has magnitude one, and $w$ has at most $k$ nonzero entries. Fourier inversion gives the required labels:
\begin{equation}
 f(X_j)=\sum_{S\in C}\widehat f(S)\chi_S(X_j)=\langle v(X_j),w\rangle.
 \label{eq:measurement}
\end{equation}
For $M\ge100$, apply Lemma~\ref{lem:cp} with $\mu=1$, $\beta=\ln20$, and budget~\eqref{eq:recovery-budget}. The true vector $w$ is the unique optimum with failure probability at most $5/100+1/20=0.1$.

For $M<100$, append $100-M$ independent uniform signs $\varepsilon$ to each row, independently of the inputs and other rows; append zeros to $w$. Writing $\widetilde v\coloneqq(v,\varepsilon)$ and $\widetilde w\coloneqq(w,0)$, independence and $\E\varepsilon=0$ give
\[
 \E[\widetilde v\widetilde v^\top]=I_{100},\qquad
 \max_i|\widetilde v_i|^2=1,\qquad
 \langle\widetilde v,\widetilde w\rangle=\langle v,w\rangle.
\]
The same application of Lemma~\ref{lem:cp} in dimension $100$ makes $\widetilde w$ uniquely optimal with probability at least $0.9$. On this event, any other original feasible vector $z$ with $\|z\|_1\le\|w\|_1$ would give a competing augmented vector $(z,0)$ of no larger norm, a contradiction. Hence $w$ is also uniquely optimal in the original problem. The padding is only for the proof and uses no additional examples.

In both cases the budget is $O(k\log M)$, since $\ln(\max\{M,100\})=O(\log M)$ under our logarithmic convention. On the success event, $w$ is feasible and $k$-sparse, so Algorithm~\ref{alg:recovery} returns its Fourier list. If $C$ is random, condition on all data used to construct it. For every resulting $C$ containing $\Sf$, the dictionary and target are fixed, while the independent recovery examples remain uniform. Applying the preceding guarantee and averaging over these conditionings proves the final assertion.
\end{proof}

\begin{algorithm}[tbp]
\caption{\textsc{Recover}$(C,k)$}
\label{alg:recovery}
\begin{algorithmic}[1]
\Require A nonempty set $C$ of distinct subsets, a sparsity bound $k$, and uniform classical examples independent of the data used to construct $C$
\State $M\gets|C|$; $q\gets q(k,M)$ from~\eqref{eq:recovery-budget}
\State Draw $q$ independent examples $(X_j,Y_j)_{j=1}^q$, where $Y_j=f(X_j)$
\State Set $Y\gets(Y_1,\ldots,Y_q)^\top$ and $A_{j,S}\gets\chi_S(X_j)$ for $j\in[q]$, $S\in C$
\State Solve $\min_{z\in\R^C}\norm z_1$ subject to $Az=Y$ exactly
\If{the constraints are infeasible} \State \Return $\fail$ \EndIf
\State Let $z$ be an exact optimal solution
\If{$|\{S\in C:z_S\ne0\}|>k$} \State \Return $\fail$ \EndIf
\State \Return $\{(S,z_S):S\in C,\ z_S\ne0\}$
\end{algorithmic}
\end{algorithm}

Algorithm~\ref{alg:recovery} may return any exact optimal solution. On the success event in Lemma~\ref{lem:recovery}, the true coefficient vector is the unique optimum, so this choice cannot affect correctness. An incorrect list is counted as failure, even if it is feasible and has at most $k$ nonzero entries. The program has rational data, and we use an exact linear-programming solver that returns a rational optimum; no further tie-breaking rule is needed.

\section{The Learning Algorithm and Its Sample Complexity}
\label{sec:algorithm}
Algorithm~\ref{alg:main} combines the support construction with exact recovery. Each classical example requested by \textsc{Recover} is obtained by measuring one quantum copy in the computational basis. The recovery copies are independent of those used for Fourier sampling. The recursive construction of $C_i$ in the pseudocode is exactly the sumset in~\eqref{eq:candidates}, with repeated subsets removed after each step.

\begin{algorithm}[tbp]
\caption{\textsc{LearnFourierSparse}$(n,k)$}
\label{alg:main}
\begin{algorithmic}[1]
\Require Independent copies of $|\psi_f\rangle$; known $n$ and $1\le k\le2^n$
\State $r\gets16\max\{1,\lceil\log k\rceil\}$
\State $m\gets\lceil rk\rceil$; $g\gets2\left\lceil\ln(100k^2)/\ln(r/4)\right\rceil+1$
\State Apply the Fourier measurement of Lemma~\ref{lem:sampling} to exactly $4gm$ copies
\If{fewer than $gm$ measurements succeed} \State \Return $\fail$ \EndIf
\State Retain the first $gm$ Fourier samples and split them, in order, into $g$ groups of size $m$
\State $C_0\gets\{0\}$
\For{$i=1,\ldots,g$}
  \State Let $T_i$ be the set of distinct subsets in group $i$
  \State $C_i\gets\{U+V:U\in C_{i-1},\ V\in T_i\cup\{0\}\}$
\EndFor
\State \Return \textsc{Recover}$(C_g,k)$ using classical examples obtained by measuring additional copies
\end{algorithmic}
\end{algorithm}

\begin{lemma}[Obtaining independent Fourier samples]
\label{lem:budget}
The Fourier stage of Algorithm~\ref{alg:main} fails with probability at most $e^{-gm/4}$. Conditional on having at least $gm$ successful measurements, its retained subsets are $gm$ independent Fourier samples.
\end{lemma}
\begin{proof}
Let $N$ count the successes among the $4gm$ measurements. By Lemma~\ref{lem:sampling}, $N$ is binomial with mean $\mu=2gm$. Apply the Chernoff--Hoeffding bound (Proposition~\ref{prop:chernoff}) with $\eta=1/2$:
\[
 \Pr[N<gm]\le\Pr[N\le\mu/2]
 \le\exp\!\left(-\frac{(1/2)^2(2gm)}2\right)
 =e^{-gm/4}.
\]

For the distributional claim, each measurement can equivalently be generated by an independent fair success flag and an independent subset drawn from $\widehat f(S)^2$, with the subset revealed only on success. This construction is valid because
\[
 \Pr[\text{success and output }S]=\tfrac12\widehat f(S)^2.
\]
The event $N\ge gm$ depends only on the flags. Conditional on any flag sequence with this property, the retained subsets come from distinct independent draws. Their joint distribution is the product Fourier distribution, independent of which successful positions are retained. Averaging over the flag sequences proves the statement.
\end{proof}

\begin{proof}[Proof of Theorem~\ref{thm:main}]
The proof has three parts: we establish constant success probability, bound the number of copies on every execution, and amplify success.

\paragraph*{Success probability.}
Conditional on sufficient Fourier outcomes, Lemma~\ref{lem:budget} allows us to apply Theorem~\ref{thm:coverage} to $C_g$. Conditional further on any first-stage transcript with $\Sf\subseteq C_g$, the copies reserved for \textsc{Recover} give independent uniform classical examples. Lemma~\ref{lem:recovery} then bounds the recovery failure by $0.1$. Taking the union bound over these three types of failure gives
\[
 \Pr[\text{incorrect output or }\fail]
 \le e^{-gm/4}+\frac1{50r}+0.1<\frac13.
\]
Indeed, $r\ge16$, $m\ge16$, and $g\ge1$, so
\[
 e^{-gm/4}+\frac1{50r}+0.1
 \le e^{-4}+\frac1{800}+\frac1{10}<\frac13.
\]
This proof includes sparsity one. On failure events the algorithm may return an incorrect list or $\fail$; neither event changes its copy bound.

\paragraph*{Sample bound.}
First suppose $k\ge3$. For a general sampling ratio $r\ge16$, the Fourier and recovery stages use $O(krg)$ and $O(kg\log k)$ copies, respectively. Taking $r=\Theta(\log k)$ balances these contributions. Algorithm~\ref{alg:main} chooses
$r=16\lceil\log k\rceil=\Theta(\log k)$.
The definitions in~\eqref{eq:parameters} therefore give
\[
 m=O(k\log k),\qquad
 g=O\!\left(\frac{\log k}{\log\log k}\right).
\]
Indeed, $\ln(r/4)=\Theta(\log\log k)$ and $\ln(100k^2)=\Theta(\log k)$; rounding only changes these bounds by constants. The Fourier stage thus uses
\[
 4gm=O\!\left(\frac{k\log^2 k}{\log\log k}\right)
\]
copies. By~\eqref{eq:candidate-size} and Lemma~\ref{lem:recovery}, the recovery stage uses at most
\[
 q(k,|C_g|)\le q(k,(k+1)^g)=O(kg\log k)
 =O\!\left(\frac{k\log^2 k}{\log\log k}\right)
\]
additional copies. The bound $|C_g|\le(k+1)^g$ holds for every measurement record, even when $C_g$ omits a subset in the true support. Each recovery example uses one additional quantum copy, so both stages have a total budget of $O(k\Lambda(k))$ on every execution.

For $k=1,2$, the chosen $r$, $m$, and $g$ are all bounded by absolute constants, as is $|C_g|$; the same algorithm uses $O(1)$ copies. This proves the sample bound in Theorem~\ref{thm:main} for the entire parameter range. Every quantum copy is measured separately, either by the Fourier measurement or in the computational basis; all subsequent processing is classical.

\paragraph*{Amplification.}
Run the learner independently an odd number $R$ of times. Write each nonfailure output in a canonical form by sorting the incidence vectors of the subsets lexicographically and reducing each rational coefficient to lowest terms. Return the list appearing in a strict majority of the runs, and return $\fail$ if there is no such list. This is a vote on complete representations, not on individual coefficients.

Let $Z_i\in\{0,1\}$ equal one exactly when run $i$ returns $\fail$ or an incorrect representation, and set $S\coloneqq\sum_{i=1}^R Z_i$. The runs use independent copies, so the $Z_i$ are independent. The success guarantee gives $\E Z_i\le1/3$ and hence $\E S\le R/3$. If $S<R/2$, a strict majority returns the same correct list. Therefore the final failure probability is at most $\Pr[S\ge R/2]$.

On the event $S\ge R/2$, we have
\[
 S-\E S\ge\frac R2-\frac R3=\frac R6.
\]
Applying Hoeffding's inequality (Proposition~\ref{prop:hoeffding}) with $t=R/6$ gives
\begin{equation}
 \Pr[S\ge R/2]
 \le \Pr[S-\E S\ge R/6]
 \le \exp\!\left(-\frac{2(R/6)^2}{R}\right)
 =e^{-R/18}.
 \label{eq:amplification}
\end{equation}
Choose the smallest odd integer $R\ge18\ln(1/\delta)$. Then $R=O(\log(1/\delta))$, and the failure probability is at most $\delta$. Multiplying the fixed budget of one run by $R$ proves the amplification claim in Theorem~\ref{thm:main}.
\end{proof}

\section{Classical Sample Complexity}
\label{sec:lower}
We prove Theorem~\ref{thm:classical} by applying sparse recovery to the full dictionary and combining the Haviv--Regev lower bound with Proposition~\ref{prop:quantum-lower}.

\begin{lemma}[Consequence of the Haviv--Regev lower bound~{\cite[Theorem~3.7]{HR}}]
\label{lem:hr}
For every $n\ge1$ and $1\le k\le2^n$, any learner using independent uniform classical examples requires $\Omega(k(n-\log k))$ examples in the worst case to identify every Boolean function $f:\F_2^n\to\{-1,1\}$ of Fourier sparsity at most $k$ exactly with probability at least $2/3$.
\end{lemma}

Haviv and Regev use $\{0,1\}$-valued targets. Every nonzero such target $h$ has $\widehat h(\varnothing)>0$, so $1-2h$ has no larger Fourier support; the zero function maps to the constant function $1$. Relabeling the examples therefore transfers their lower bound to our sign convention. For arbitrary $k\ge2$, apply their construction at $K=2^{\lfloor\log k\rfloor}$: $K>k/2$ and $n-\log K\ge n-\log k$. For $k=1$, signed parity functions give the required $\Omega(n)$ bound~\cite{ACLPW}.

\begin{proof}[Proof of Theorem~\ref{thm:classical}]
For the upper bound, take $C=\F_2^n$ in Lemma~\ref{lem:recovery}. This set always contains $\Sf$, so $q(k,2^n)=O(kn)$ uniform classical examples suffice.

For the lower bound, Lemma~\ref{lem:hr} gives $\Omega(k(n-\log k))$. A quantum learner can measure each copy in the computational basis to obtain one classical example. Hence the quantum lower bound in Proposition~\ref{prop:quantum-lower} also applies to classical samples and gives the additional bound $\Omega(k\log k)$. Both bounds must hold, and
\[
 \max\{k(n-\log k),k\log k\}\ge\frac{kn}{2}.
\]
Thus the classical sample complexity is $\Omega(kn)$, matching the upper bound.
\end{proof}

\section*{Acknowledgements}
The author was supported by start-up research funding from Fuzhou University (Grant No.\ XRC-26004).

\paragraph*{Disclosure of Generative AI Use.}
ChatGPT (model: GPT-6 Astra) was used in developing the arguments and preparing this manuscript. Its uses included exploring candidate-set constructions, expanding intermediate proof steps, checking the quantum lower-bound reduction and the applicability of cited recovery results, searching for related literature, organizing the exposition, and drafting text. The author is responsible for the mathematical claims, references, and final presentation.

\clearpage
\phantomsection\label{page:references}
\bibliography{references}

\clearpage
\appendix
\section{A Quantum Lower Bound of Sample Complexity}
\label{app:quantum-lower}\label{app:balanced-conversion}\label{page:appendix-start}
Theorem~5 of~\cite{ACLPW} establishes $\Omega(k\log k)$ uniform quantum examples for $k\le2^{cn}$ and fixed $c\in(0,1)$. The following all-parameter version is a consequence of the quantum coupon-collector theorem~\cite{QCC}.

\begin{proposition}[Quantum lower bound]
\label{prop:quantum-lower}
For every $n\ge1$ and $1\le k\le2^n$, any quantum learner requires $\Omega(k\log k)$ uniform quantum examples in the worst case to identify every Boolean function $f:\F_2^n\to\{-1,1\}$ of Fourier sparsity at most $k$ exactly with probability at least $2/3$. The bound holds even for learners using arbitrary joint measurements.
\end{proposition}

We use the following specialization of the quantum coupon-collector lower bound.

\begin{lemma}[Quantum coupon collection~{\cite[Theorem~6]{QCC}}]
\label{lem:qcc}
Let $K\ge4$ be even, and let $A\subseteq[K]$ be an unknown set of size $K/2$. Given independent copies of
\[
 |A\rangle\coloneqq\sqrt{\frac2K}\sum_{i\in A}|i\rangle,
\]
any procedure requires $\Omega(K\log K)$ copies in the worst case to identify every such $A$ with probability at least $2/3$. The bound holds even for procedures using arbitrary joint measurements.
\end{lemma}
The cited theorem concerns an unknown $a$-element subset of an $N$-element universe and gives $\Omega(a\log\min\{a,N-a\})$ copies. Taking $N=K$ and $a=K/2$ yields the statement above. Only copies of $|A\rangle$ are supplied; no oracle depending on $A$ is available.

\begin{lemma}[Converting a subset state into a labeled example]
\label{lem:balanced-conversion}
For every even $K\ge2$, there is a unitary $U_K$, independent of $A$, such that every $A\subseteq[K]$ of size $K/2$ satisfies
\begin{equation}
 U_K\bigl(|A\rangle\otimes|+\rangle\bigr)
 =|\psi_A\rangle,
 \qquad
 |\psi_A\rangle\coloneqq\frac1{\sqrt K}\sum_{i=1}^K|i\rangle|\mathbf1_A(i)\rangle.
 \label{eq:balanced-conversion}
\end{equation}
Here $\mathbf1_A(i)$ equals $1$ if $i\in A$ and $0$ otherwise. Thus one copy of $|A\rangle$ suffices to prepare exactly one uniform quantum example for its membership function.
\end{lemma}
\begin{proof}[Proof of Lemma~\ref{lem:balanced-conversion}]
For $\bar A\coloneqq[K]\setminus A$, define
\[
 |\bar A\rangle\coloneqq\sqrt{\frac2K}\sum_{i\notin A}|i\rangle,
 \qquad |u_K\rangle\coloneqq\frac1{\sqrt K}\sum_{i=1}^K|i\rangle,
 \qquad D_K\coloneqq2|u_K\rangle\langle u_K|-I_K.
\]
The reflection $D_K$ fixes $|u_K\rangle$ and negates orthogonal vectors, so it is unitary. Since $|A|=|\bar A|=K/2$, we have $\langle u_K|A\rangle=1/\sqrt2$ and $\sqrt2|u_K\rangle=|A\rangle+|\bar A\rangle$. Hence
\[
 D_K|A\rangle=\sqrt2|u_K\rangle-|A\rangle=|\bar A\rangle.
\]
Append $|+\rangle$ and apply the controlled unitary
\[
 U_K\coloneqq D_K\otimes|0\rangle\langle0|+I_K\otimes|1\rangle\langle1|,
\]
which is unitary on each orthogonal label subspace. Then
\begin{align*}
 U_K(|A\rangle|+\rangle)
 &=\frac{|\bar A\rangle|0\rangle+|A\rangle|1\rangle}{\sqrt2}\\
 &=\frac1{\sqrt K}\left(\sum_{i\notin A}|i,0\rangle+\sum_{i\in A}|i,1\rangle\right)=|\psi_A\rangle.
\end{align*}
All operations depend only on $K$, not on $A$: no reflection about the unknown state, membership query, or postselection is used.
\end{proof}

\begin{proof}[Proof of Proposition~\ref{prop:quantum-lower}]
First suppose $k\ge4$. Set $d\coloneqq\lfloor\log k\rfloor$ and $K\coloneqq2^d$, so $2\le d\le n$ and $k/2<K\le k$. For $x=(x_1,\ldots,x_n)$, define
\[
 i(x)\coloneqq1+\sum_{j=1}^d2^{j-1}x_j\in[K].
\]
For every $A\subseteq[K]$ of size $K/2$, let
\[
 f_A(x)\coloneqq1-2\mathbf1_A(i(x)).
\]
Thus $\mathbf1_A(i(x))$ equals $1$ precisely when $i(x)\in A$, and $f_A(x)$ is $-1$ in that case and $1$ otherwise. This function depends only on the first $d$ variables. To check its sparsity explicitly, if $S$ contains some coordinate $j>d$, pairing inputs that differ only in coordinate $j$ cancels the sum defining $\widehat f_A(S)$. Hence the Fourier support is contained in $\{S:S\subseteq[d]\}$ and has size at most $2^d=K\le k$.

Suppose a learner uses at most $T$ uniform quantum examples to learn every target of sparsity at most $k$. Starting with $T$ copies of $|A\rangle$, apply Lemma~\ref{lem:balanced-conversion} separately to each copy. Relabel $i\in[K]$ by the $d$ binary digits of $i-1$, and append $n-d$ qubits in $|+\rangle$. After ordering the registers, each resulting state is exactly
\[
 2^{-n/2}\sum_{x\in\F_2^n}|x\rangle|\mathbf1_A(i(x))\rangle
 =|\psi_{f_A}\rangle.
\]
No further unknown-state copies are used. Feed these $T$ states to the learner. On its success event, evaluate its exact classical description on the inputs $(x_1,\ldots,x_d,0,\ldots,0)$; those giving value $-1$ identify precisely the members of $A$. This evaluation needs no additional samples. If the learner returns $\fail$, or these evaluations do not yield a set of size $K/2$, return any fixed set of that size. Otherwise return the set obtained; on the learner's success event it is exactly $A$.

We have therefore obtained a quantum coupon-collection algorithm with the same copy bound and at least the same success probability. Lemma~\ref{lem:qcc} implies
\[
 T=\Omega(K\log K)=\Omega(k\log k),
\]
since $K>k/2$ and $\lfloor\log k\rfloor\ge\tfrac12\log k$. The reduction preserves the learner's arbitrary joint measurements and adaptive processing. For $k=1,2,3$, distinguishing the two constant functions requires at least one copy, which gives the asserted bound under our convention for logarithmic factors.
\end{proof}
\end{document}